\documentclass[conference]{IEEEtran}
\IEEEoverridecommandlockouts
\usepackage{cite}
\usepackage{amsmath,amssymb,amsfonts}
\usepackage{amsmath}
\usepackage{graphicx}
\usepackage{textcomp}
\usepackage{xcolor}
\usepackage{enumitem}
\usepackage{subcaption}
\usepackage[linesnumbered,ruled,vlined]{algorithm2e}

\def\BibTeX{{\rm B\kern-.05em{\sc i\kern-.025em b}\kern-.08em
    T\kern-.1667em\lower.7ex\hbox{E}\kern-.125emX}}
\begin{document}


\title{Content-defined Merkle Trees for Efficient Container Delivery}

\author{\IEEEauthorblockN{Yuta Nakamura, Raza Ahmad, Tanu Malik}
\IEEEauthorblockA{\textit{School of Computing} \\
\textit{DePaul University}\\
Chicago, IL USA \\
ynakamu1, raza.ahmad, tanu.malik@depaul.edu}
}

\newcommand{\CDMT}{\texttt{CDMT}}

\maketitle

\begin{abstract}
Containerization simplifies the sharing and deployment of applications when environments change in the software delivery chain. To deploy an application, container delivery methods push and pull container images. These methods operate on file and layer (set of files) granularity, and introduce redundant data within a container. Several container operations such as upgrading, installing, and maintaining become inefficient, because of copying and provisioning of redundant data. 


In this paper, we reestablish recent results that block-level deduplication reduces the size of individual containers, by verifying the result using content-defined chunking. Block-level deduplication, however, does not improve the efficiency of push/pull operations which must determine the specific blocks to transfer. We introduce a content-defined Merkle Tree (\CDMT{}) over deduplicated storage in a container. \CDMT{} indexes deduplicated blocks and determines changes to blocks in logarithmic time on the client. \CDMT{} efficiently pushes and pulls container images from a registry, especially as containers are upgraded and (re-)provisioned on a client. We also describe how a registry can efficiently maintain the \CDMT{} index as new image versions are pushed. We show the scalability of \CDMT{} over Merkle Trees in terms of disk and network I/O savings using 15 container images and 233 image versions from Docker Hub.

\end{abstract}


\section{Introduction}
\label{sec:introduction}
Containers are lightweight alternatives to virtual machines and are increasingly used for sharing and deploying applications. Applications are containerized by using primitives in the Linux kernel, which partition and isolate kernel resources such as process identifiers, file names, user identifiers, and hostnames. Container engines such as Docker~\cite{Docker} and Singularity~\cite{kurtzer2017singularity} make it easy to build, deliver, and run such containerized applications. Thus, when a container is built on a target machine, it can be re-executed in isolation using encapsulated data and dependencies without the target environment interfering with the computation.

Containerization improves portability of an application in a software delivery chain. But before a container runs, the system must encapsulate and provision all application data, including a Linux distribution, system dependencies, binaries, scripts, etc. The operating system downloads and allocates files on the target file system increasing container provisioning and start-up times. Docker build-time is indeed the most time-consuming step in running a containerized application~\cite{verma2015large}, \cite{lagar2009snowflock}.



To improve the storage explosion within containers, current container engines optimize the use of file-system resource. By default, engines identify redundancy at the granularity of files. If a file and its path remains same, the engine adds no new file. However, if any part of a file is modified or its path is changed, the entire file is physically copied to produce a new file. A file-granularity scheme works well as long as most modified files are small, or not significantly larger than a disk block (4KB)~\cite{wu2015totalcow}. However, for large files, identifying redundancy at the file-granularity level still causes a storage explosion. 

Block-based storage has emerged as a viable solution for the storage explosion problem in containers in which blocks instead of entire files are deduplicated~\cite{harter2016slacker}. To maintain a competitive deduplication ratio, block-based storage must distinguish between existent and new blocks, especially as images are pulled, and modified images pushed. For example, if a container registry hosts a TensorFlow image, and a client actively develops it, and changes a few files, then the client must ship only those blocks corresponding to those changes to the registry. Clearly, if the client ships the entire files (the current default), there will be an increase in provisioning and build times.  

Mapping from blocks to images and vice versa requires hash-based comparisons via an index to determine which blocks to pull or push and thus reduce network and disk overhead. Merkle trees~\cite{10.1007/3-540-48184-2_32} reduce the number of hash comparisons by composing hash values over a set of blocks. Given a Merkle tree index on two versions of an image, while the compositional hashes can detect that two versions are changed, they cannot precisely determine which blocks have changed across the two container image versions.

In this paper, we introduce Content-Defined Merkle Tree (\CDMT{}), a hash-based tree index constructed over content-defined blocks in a deduplication system. In \CDMT{}, internal nodes are also content-defined to avoid significant changes in hash values as a few blocks are modified or added over time. Intuitively, \CDMT{} has the same advantage over Merkle tree, which content-defined chunking\footnote{A chunk is synonymous to a block} has over fixed length chunking. Fixed length chunking suffers from the shifted content problem, since a single byte insertion or deletion completely changes the  boundaries of other chunks and thus their hash values. Content-defined chunking, which creates chunk boundaries based on patterns in content and not fixed length, avoids this byte-shift problem. Similarly, hashes of internal nodes in Merkle trees change entirely due to block insertions, but \CDMT{}, being content-defined, structurally remains resistant to this change. 


There are several advantages of using a content-defined tree index over a Merkle tree index. Given two image versions, a Merkle tree comparison might conclude that the entire container image has changed when only a single block may have changed. On the contrary, a \CDMT{} index detects the precise blocks changed between two image versions. If the versions are resident on a client and a registry, then only the changed blocks are exchanged with a \CDMT{} index, and, as our experiments show, this precise exchange translates to smaller network footprint. The \CDMT{} index over content-defined blocks is useful when data-intensive container images are part of the delivery chain: small updates on a large-sized file can require current container engines to duplicate an entire layer or image version. Typically, an index is much smaller (in our case $\thicksim$Kbs) than data chunks and is thus memory-efficient.  Since the chunks are deduplicated and not compressed, the content-defined index continues to provide disk I/O savings.



\CDMT{} is write-optimized in that when it indexes new blocks, the index itself need not be reorganized. \CDMT{}  versions internal nodes using node-copying and organizes the pointers such that a given version of the container image can be obtained in linear time. Using node-copying, rebuilding container images during continuous delivery with small updates is much faster than reprovisioning the entire image. By maintaining versions within the \CDMT{} index, the tree serves both as an index over chunks and as an authentication mechanism to reconstruct the container image correctly.

The contributions of this paper are as follows:
\begin{itemize}
\item We use content-defined chunking for deduplicating blocks, and show that CDC-based block deduplication reduces overall storage size when storing a large number of images. We compare our CDC-based deduplicated store against simply storing compressed images---a method commonly used in container registries for pushing and pulling images. We also compare it with uncompressed images within a container which uses union file systems. We show an improvement of 8x over compressed images. 
\item We introduce content-defined Merkle trees (\CDMT{}) and show how they compare image (or image layer) versions, and detect changes to relevant blocks. \CDMT{}  improves deduplication ratio by 10-15\% across versions of images.
\item We develop algorithms for efficiently maintaining content-defined Merkle tree (\CDMT{}) using node-copying method such that a specific application version is obtained in time linear in the number of versions.
\item We describe how the block-based deduplication and it's accompanying index is used to push and pull specific image versions. 
\item We conduct a thorough experimental analysis on 15 images and 233 image versions from Docker Hub. Experiments show that deduplication reduces total size and that the deduplication storage when supported with a content-defined Merkle tree index  efficiently push/pull layers over the network otherwise network communication of chunks increase by over 40\%.
\end{itemize}
We organize the rest of the paper as follows: Section~\ref{sec:rw} describes related work. Section~\ref{sec:bg} provides background on content-defined chunking (CDC) and Merkle trees, and describes the chunk-shift problem in Merkle trees. 
Section~\ref{sec:cdmt} describes an algorithm for creating  content-defined Merkle trees (\CDMT{}), and analyzes the algorithm. Section~\ref{sec:system} considers how to build \CDMT{} index in container engines, and maintain them across clients and container registries, especially as new versions are created. 
We describe implementation and experiments in Section~\ref{sec:experiments}, and  conclude in Section~\ref{sec:conclusion}.


\section{Related Work}
\label{sec:rw}
The problem of efficiently maintaining encapsulated data in containers has received significant attention. In this context, we review storage mechanisms within containers and efficient maintenance of deduplicated content. 


Harter \textit{et.} \textit{al.}~\cite{harter2016slacker} show that individual containers are ``bloated’’ due to isolation \textit{i.e.,} breaking down an application into many smaller processes, each of which when isolated leads to loading nearly the same set of libraries, and thus duplication of data. They also show that containers utilize fewer data than they contain, \textit{i.e.,} containers package software with more dependencies (to use in a variety of environments and by a range of applications), while in practice containers utilize only a limited subset. To counter bloating, Docker~\cite{Docker} uses compression methods while transporting image layers. These layers, however, must be uncompressed before use thus increasing the overall size of containers.  

A recent study~\cite{verma2015large} shows that, even with compression, containers involve significant copying and installation overheads with package installation taking about 80\% of the total time. 
Compression is efficient for small sized containers but registries (large-scale aggregation of containers) or a large container with several layers will benefit from block-level deduplication as posited in~\cite{harter2016slacker}. In this work, we have  considered content-defined chunking as a method for block-level deduplication. 



Destor\cite{zhang2016fast} shows the efficiency of a deduplication system critically depends on the size of a lookup index, which determines if a block exists in the deduplicated system. To the best of our knowledge, ours is the first paper considering an index over block-level deduplication in container systems, and assessing the impact of an index on push/pull operations. Systems like Picky~\cite{hintze2016picky} have used Merkle tree for applying patches and upgrades to published datasets. Merkle trees are used for hash-based comparisons in cryptoanalysis~\cite{munoz2004certificate}, version control~\cite{vaidya2019commit} and block chain systems~\cite{zhang2018towards}. We show that Merkle trees are not sufficient as an indexing structure over block-based deduplication system.



\cite{tarasov17} explores different Docker storage architectures  and how a storage solution impacts system
and workload performance. They experimentally show Btrfs~\cite{rodeh2013btrfs} as being more space and I/O efficient over AUFS and union file systems for the same number of Docker layers. While we have not implemented \CDMT{} within a specific storage architecture, \CDMT{} relies on copy-on-write functionality available 
as part of the underlying filesystem.
If a copy-on-write block is determined, \CDMT{} creates internal node copies for that path. 

\section{Preliminaries}
\label{sec:bg}
We describe the issue of using Merkle trees within block deduplicated storage. For this we briefly describe Merkle trees, and content-defined chunking, which forms the basis for block deduplicated storage. 

\subsection{Content-defined chunking} 
Content-defined chunking (CDC) is a subfile-level deduplication method that aims to detect redundancy within files~\cite{Muthitacharoen_ContentBaseddeDuplication}. Without the CDC, a deduplication system can partition files into fixed-width chunks, and use resulting chunk hashes to detect identical chunks across two files. A fixed-width chunking-based deduplication system suffers from the ``byte-shift'' problem in which insertion of a single byte at the beginning of the file changes the fixed-width partition boundaries, and thus the hash values of all the chunks, even though the data in the remainder of the file has not changed. As a result, fixed-width chunking-based deduplication system suffers from low deduplication ratio. Content-defined chunking (CDC) aims to detect redundancy by creating variable-length chunks based on  file content and uses hash fingerprint of variable length chunks to detect common chunks. Chunks are created if the chunk content matches a predefined pattern such as 5 least significant bytes of the hash are 0. If the content matches the pattern, the chunk is `cut' and its boundary defined.  
CDC solves the problem of byte-shift because even if the byte is inserted at the beginning of the file, as long as the boundary pattern remains the same, the boundary of chunks is detected, and most chunks remain unchanged (Since the rest of the file content has not changed content-defined boundary patterns are unchanged). CDC uses a rolling hash method~\cite{Rabin:fingerprinting} to compute the hash values of variable-length chunks, which enable the content-based hash to be computed in time linear to the size of the content.

\subsection{Merkle tree}
The Merkle tree is a complete $k$-ary tree in which values of internal nodes are one-way hash functions of the values of their children's hash value~\cite{10.1007/3-540-48184-2_32}. Thus, leaf values are hash values of leaf's content, and Merkle tree computes the hash identifier of an internal node by applying a one-way hash function on the concatenation of hash values of its $k$ children. 

\begin{figure}[t]
\centering
\includegraphics[width=2.5in]{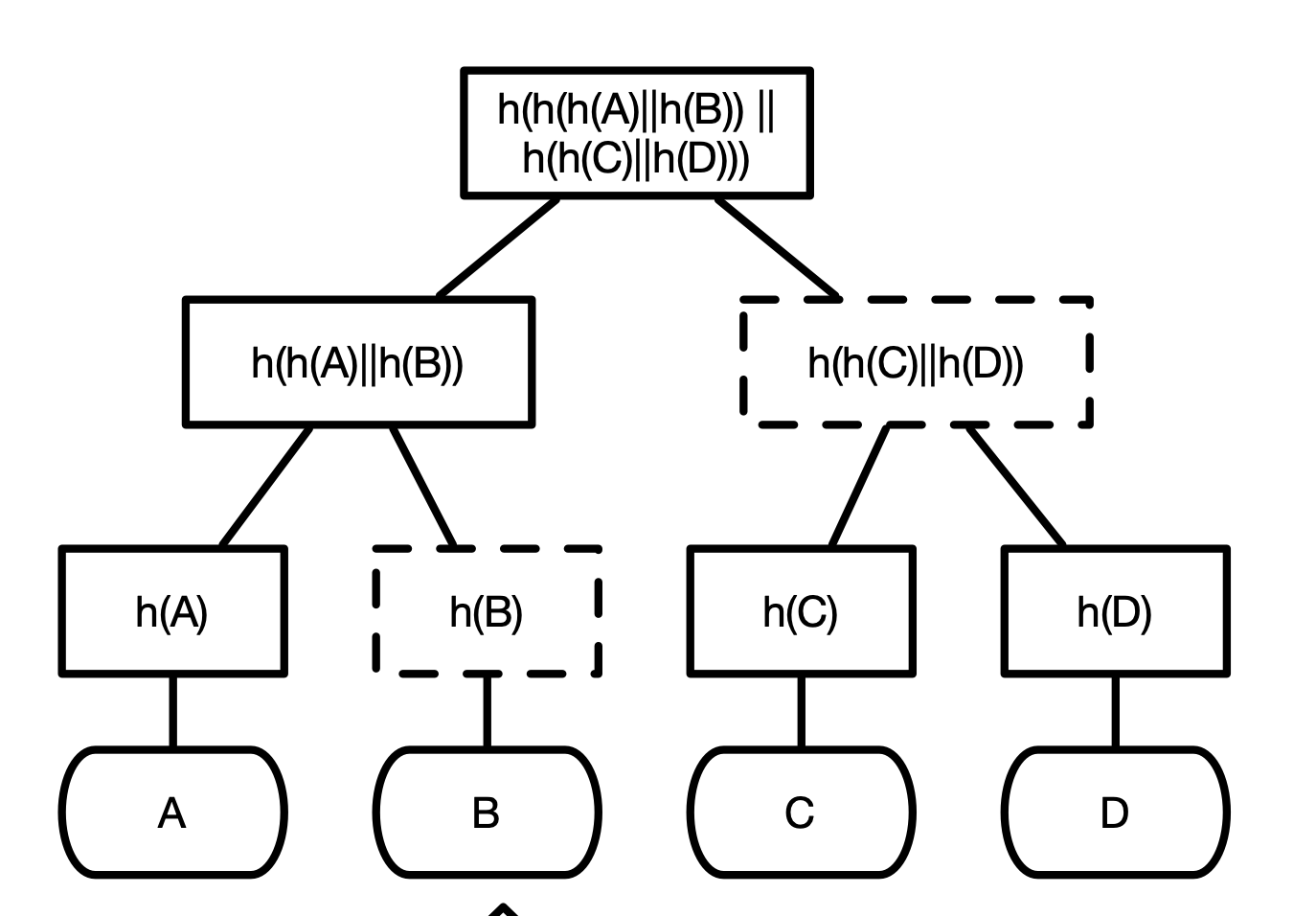}
\caption{A binary Merkle tree. The authentication path for leaf at index 0 is shown in dashed nodes}
\label{fig:merkletree}
\end{figure}

If any of the leaf nodes change then hash values from the node to the root change. Version control systems use this property of Merkle trees to determine if leaf data nodes have changed. To determine which specific leaf node has changed, two Merkle trees are compared using an \textit{authentication path}~\cite{10.1007/3-540-48184-2_32}. Figure~\ref{fig:merkletree} shows the authentication path (dotted nodes) for the leaf `A' in a binary Merkle tree. The authentication path of leaf `A' consists of the siblings of all nodes on the path from this leaf to the root. Given an original Merkle tree, and a changed Merkle tree, if the authentication path generates the new root then the corresponding leaf data must be added. Authentication paths reduce the number of hash nodes necessary for comparison~\cite{10.1007/3-540-48184-2_32}. Merkle trees determine if leaf nodes have changed in $O(\log N)$ time, where $N$ represents the number of leaf nodes, which is more efficient than $O(N)$ key-value lookup over all leaves. 





\subsection{The chunk-shift problem in Merkle trees}
\label{sec:problem}



\begin{figure}[ht]
\begin{subfigure}{0.5\textwidth}
  \centering
  \includegraphics[width=\linewidth]{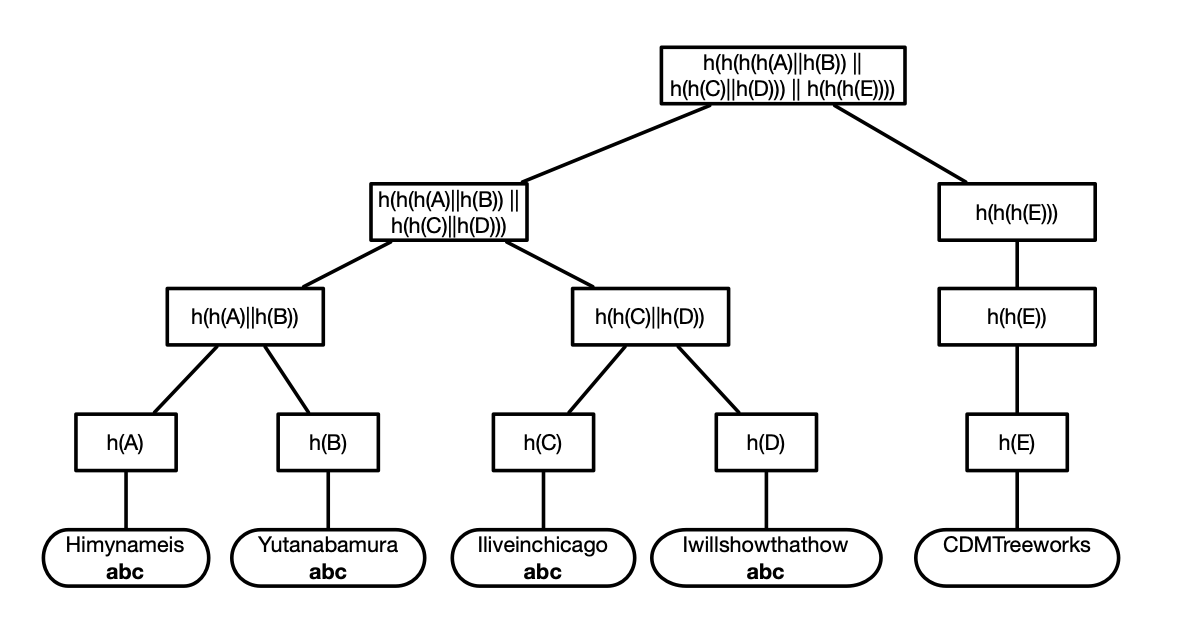}  
  \caption{A binary Merkle tree on 5 CDC chunks delimited with chunk boundary `abc'.}
  \label{fig:original}
\end{subfigure}
\begin{subfigure}{.5\textwidth}
\vspace{5pt}
  \centering
  \includegraphics[width=\linewidth]{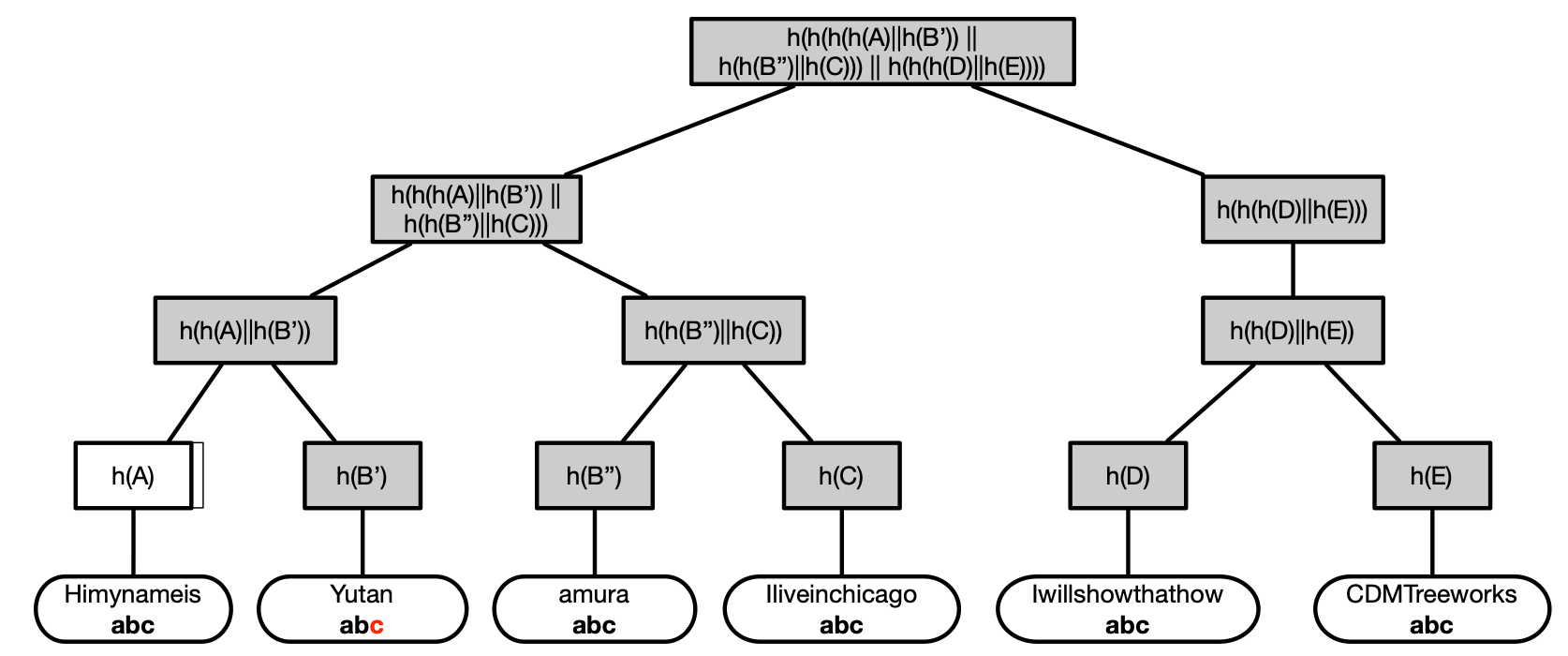}  
  \caption{The second chunk in Figure~\ref{fig:original} after the insertion of `c' splits into two due to  chunk boundary rule. The new chunk on content ``amura\textbf{abc}" shifts the hash values and changes the previous Merkle tree almost entirely.}
  \label{fig:inserted}
\end{subfigure}
\begin{subfigure}{0.5\textwidth}
\vspace{1pt}
  \centering
  \includegraphics[width=\linewidth]{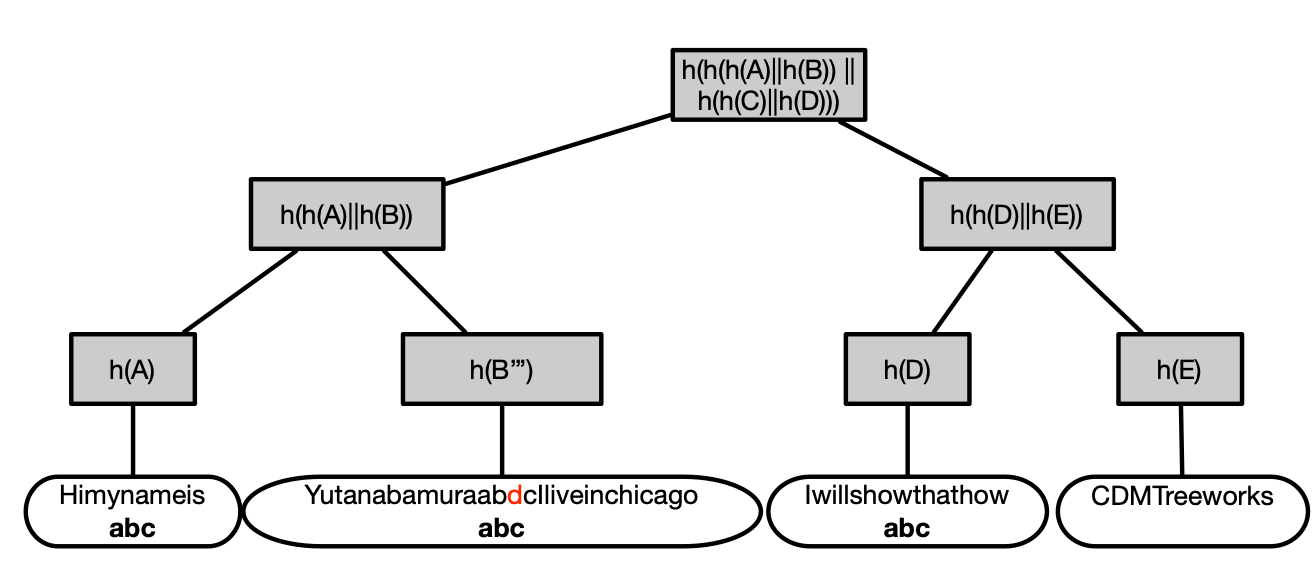}  
  \caption{The Merkle tree also changes entirely if the boundary rule is modified. Figure shows the Merkle tree when `d' is inserted within the boundary rule of second chunk of Figure~\ref{fig:original}.}
  \label{fig:deleted}
\end{subfigure}
\caption{The chunk-shift in Merkle tree over CDC chunks}
\label{fig:MTshift}
\end{figure}

Merkle trees suffer from a `chunk shift' problem on CDC-defined variable-length chunks similar to the `byte-shift' problem that exists if fixed-length chunks are deduplicated. Consider a Merkle tree on variable-length chunks obtained by CDC as shown in Figure~\ref{fig:original}. In this example, for simplicity, we have used a readable pattern of \textit{``abc"} as the pattern for CDC to demarcate a chunk boundary. 

The simpler case is when a chunk is modified without changing the chunk boundary pattern. In this case, the hash values of the node to the root are modified but there is no change in the number of CDC chunks. The hash value of most internal nodes remains same. Given two Merkle trees, one in which the chunk is modified, and one in which it is not, the modified chunks can be determined by comparing the hash values in the authentication paths of modified chunks with the original hash values.


A chunk-shift arises when the chunk boundary pattern is itself modified. In this case new chunk blocks are  either created or merged. If the number of underlying chunks change, the entire Merkle tree is modified, including hash values of internal nodes, and the height of the tree. In Figure~\ref{fig:MTshift}, the second chunk is modified by disrupting the boundary pattern and inserting the character `c' after ``yutanab''. CDC splits  the chunk into two chunks: one with ``yutanabc" and the other with ``amuraabc". Figure~\ref{fig:inserted} shows the resulting Merkle tree on the new set of chunks.
The change in Merkle tree is shown as grayed boxes, and includes the previous third, fourth and fifth chunks: their hash values h(C), h(D), and h(E) are same but since the child order position is changed the internal nodes are not found in the authentication path. Hash values in an original and modified Merkle trees, in this case, show all chunks as changed even though large number of chunks are still the same. This mismatch happens due to shift in Merkle tree hash values, which itself occurs due to a shift in the number of chunks. If nodes are merged due to an elision of pattern, a dissimilar Merkle tree is constructed, which falsely claims all chunk nodes as changed. Figure~\ref{fig:deleted} changes the pattern in the second chunk to only ``ab" by elision of a `c' thus merging the second and the third chunk. When chunks are merged, as the example shows, the height of the Merkle Tree changes, and all leaves of Figure~\ref{fig:deleted} are 
considered modified. 

Chunk shift problem defeats the purpose of Merkle tree in  because all the internal nodes from the right of the chunk shift changes, and in the worst case scenario, the tree height can change. We show in Experiments (Section~\ref{sec:experiments}) that changes to boundary patterns are common, and so are chunk shifts. 
We now describe the design of content-defined Merkle tree which is resilient to the chunk shift problem.

\section{Content-defined Merkle Tree}
\label{sec:cdmt}

The key idea in the \CDMT{} index is to construct a Merkle tree in which the internal nodes are robust to chunk shifts. This is achieved by constructing internal nodes with a variable number of children such that their hash value matches a predefined pattern. 

To match the predefined pattern, similar to CDC, \CDMT computes a a rolling hash over a fixed window consisting of children hashes, \textit{i.e.,} initially, a fixed window of $m$ children are assigned to an internal node. If the combined hash values of all the $m$ children in a window do not meet a predefined pattern, an additional child is assigned to the internal node, and the fixed window is rolled over to comprise of the new child and $m-1$ previous children and check if the new window hash meets a predefined pattern. Constructing internal nodes of a hash index based on a pattern keeps the tree robust to chunk shifts. We do not construct a formal proof for this robustness but show it experimentally and through an example below. A \CDMT{} is not balanced as a $k$-ary Merkle tree, but owing to hashes being random, it has low height, which keeps it efficient for traversal. 

\begin{figure}[t]
\begin{subfigure}{0.5\textwidth}
  \centering
  \includegraphics[width=\linewidth]{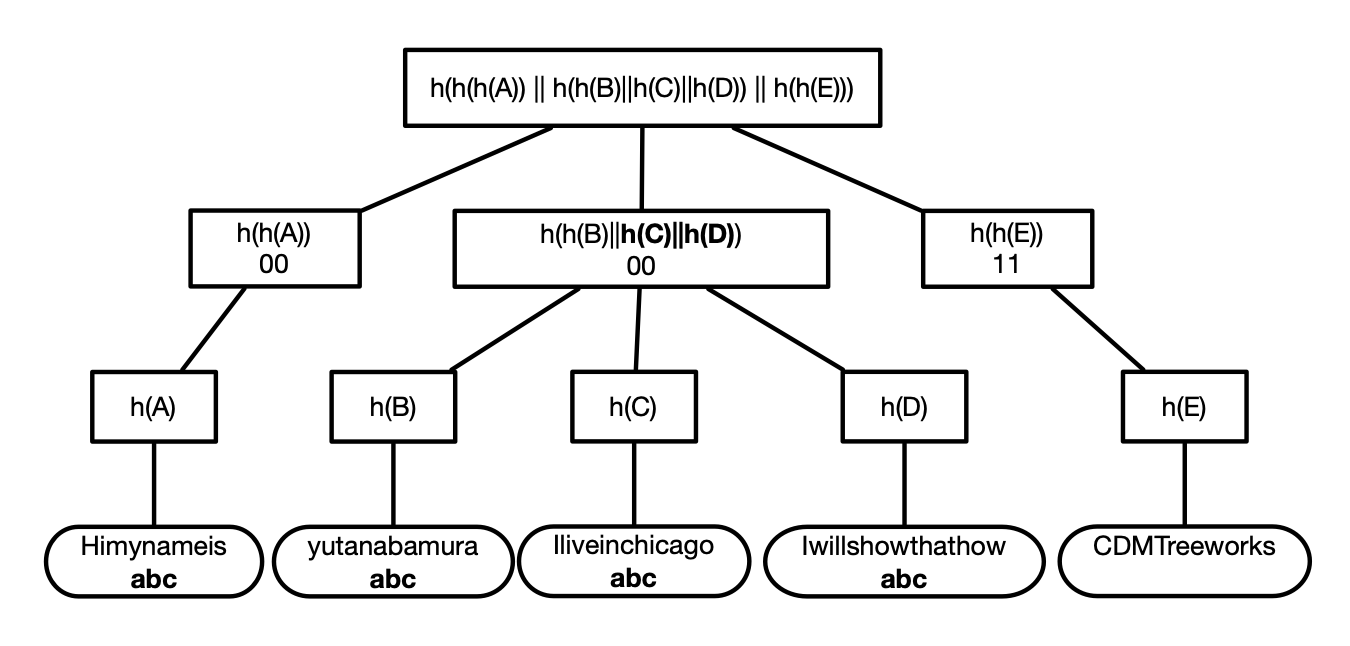}  
  \caption{A \CDMT{} tree on 5 chunks delimited with chunk boundary \textit{``abc"}. Bold hash values show hashes contributing to parent boundary.}
  \label{fig:CDMT}
\end{subfigure}
\begin{subfigure}{.5\textwidth}
\vspace{10pt}
  \centering
  \includegraphics[width=\linewidth]{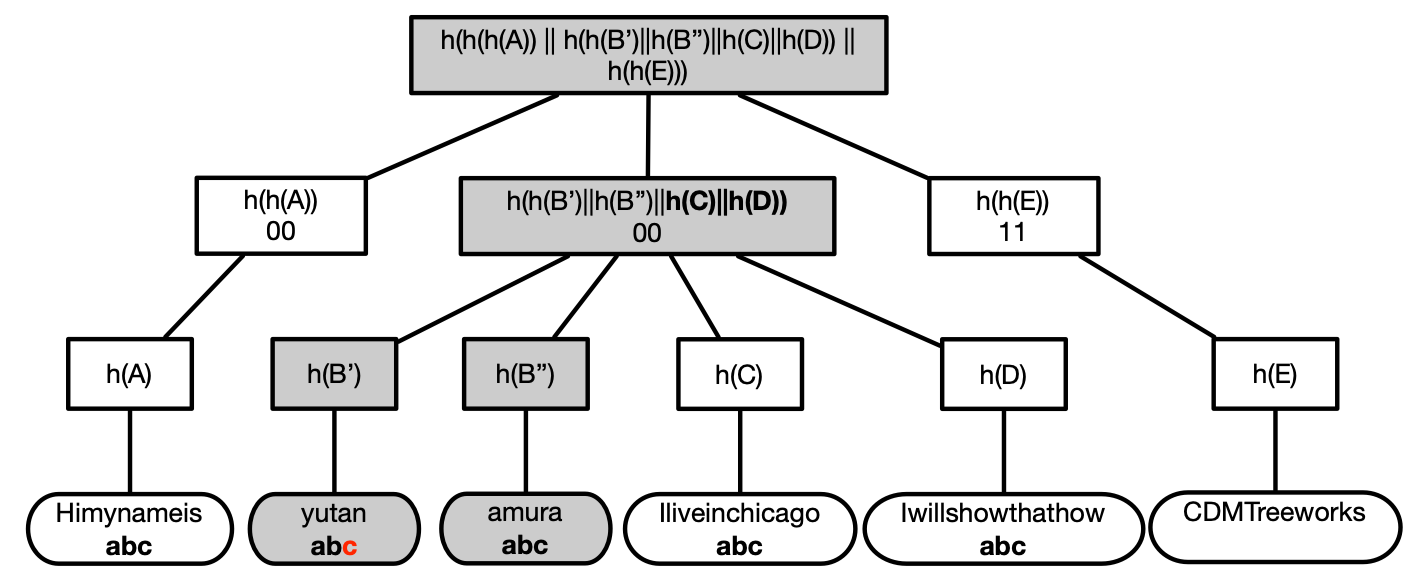}  
  \caption{The second chunk after the insertion of `c' splits into two due to  chunk boundary rule. The new chunk's changes are localized as parent node is only determined by h(C) and h(D).}
  \label{fig:CDMTinserted}
\end{subfigure}
\begin{subfigure}{0.5\textwidth}
\vspace{8pt}
  \centering
  \includegraphics[width=\linewidth]{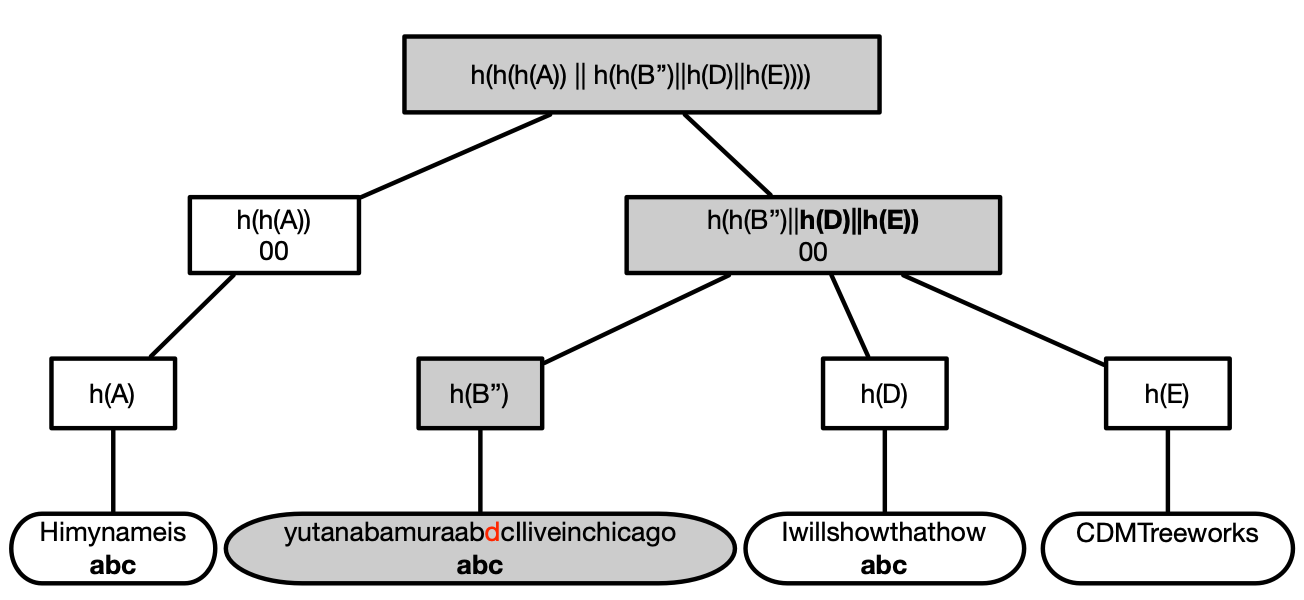}  
  \caption{The \CDMT{} tree keeps the changes localized when `d' is inserted and the boundary rule of second chunk merges the parents but does not cause a chunk shift.}
  \label{fig:CDMTdeleted}
\end{subfigure}
\caption{The shift in \CDMT{} tree over CDC chunks results in localized changes to parent nodes. In this example, for illustration, chunk boundaries (\textit{``abc"}) are different from internal node boundaries (`00'). The window size is assume to be two.} 
\label{fig:CDMTshift} 
\end{figure}

We illustrate the robustness of \CDMT{} to chunk shift through an example (Figure~\ref{fig:CDMTshift}). Consider the data chunks as previously in Figure~\ref{fig:original}, with the same data chunk boundary condition of \textit{``abc"}. Since the content of internal nodes are hash values, we consider a boundary forming rule if the last 2 bits of the hash values are 0, \textit{i.e.}, 00. A resulting \CDMT{} using Blake-2b~\cite{blake2b} hash function is shown in Figure~\ref{fig:CDMT}. In this Figure, the internal nodes are not defined based on the hash value of the concatenation of hash values of child nodes, but they are defined if the rolling hash value of the concatenation of child hash values matches the boundary rule of internal nodes. 

Applying CDC-like rule to internal nodes keeps the \CDMT{} index robust against chunk shifts. This is shown in Figure~\ref{fig:CDMTinserted}, in which even after the insertion of `c' in ``yutanabamura", the chunk is split into two and the tree structure as a whole is preserved with only the path from the new chunk `umura' to root node changing (shown in shade). This is because the rolling hash still splits after h(C) and h(D). Similarly when `d' is inserted such that the insertion breaks the chunk boundary rule (Figure~\ref{fig:CDMTdeleted}), in the \CDMT{} index the two child nodes are merged to the left but the other nodes  of the tree are resistant to the shift as the child hash values appear in the same order.
The efficiency of the \CDMT{} index depends upon an appropriately chosen window size, which determines the number of child hashes whose concatenation must be checked against the pattern rule. In our example, we have assumed a window size of 2 for internal nodes. Alternatively, if all children are chosen then the hash changes every time a child changes.
As we will show in Section~\ref{sec:experiments} the \CDMT{} index performs well with a window size of 8 where updates are provided by new versions of chunks.


Algorithm~\ref{algo:cdmt_build} show the pseudo-code for building the \CDMT{} index. The algorithm takes as input a list of hashids of content-defined chunks $L$, a pattern matching rule $R$, and a window-size of $W$. We assume the Blake-2b as our hashing function, which provides maximum efficiency for large-sized datasets. The algorithm relies on a hashmap $hm$ for quick lookups. The keys of the hashmap is the hash of the data for leaf nodes and hash for concatenated hashes of children for the internal nodes (root node included). Values of the hashmap are the pointers to the nodes in the memory. The Algorithm maintains the tree as a queue. The algorithm creates new internal nodes based on concatenation of hashids of $w$ children (Line 14 and 20). If the hash of the internal node matches the content-defined rule (Line 17), then the Algorithm declares a new parent and inserts into the queue till it finds a new root.



\begin{algorithm}
\SetAlgoLined
\DontPrintSemicolon
\SetKwProg{Def}{}{:}{}
\SetKwInOut{Input}{Input}\SetKwInOut{Output}{Output}
\SetKwData{Rule}{chunking\_rule}
\SetKwData{WindowSize}{window\_size}
\SetKwData{Queue}{S}
\SetKwData{Queue}{T}
\SetKwFunction{HF}{HashFunction}
\SetKwFunction{Insert}{insert}
\SetKwFunction{Push}{push}
\SetKwFunction{Pop}{pop}
\SetKwFunction{Size}{size}
\SetKwFunction{Copy}{Copy}
\SetKwFunction{NumChildreninWindow}{NumChildreninWindow}
\SetKwFunction{Children}{Children}
\SetKwFunction{AddChild}{AddChild}
\SetKwFunction{AddinWindow}{AddinWindow}
\SetKwFunction{NewParent}{NewParent}
\SetKwFunction{RollWindow}{RollWindow}
\Input{Queue $S$ of CDC hashids, Rule $R$, WindowSize $W$}
\Output{$T$ a \CDMT{} Tree}
\BlankLine
List $T1 = T2 \leftarrow$ null\;
Tree $T \leftarrow$ null\;
Hashmap $hm$ $\leftarrow$ null\;
\While{$S$.\Size{} $>$ 0}{
$s = S$.\Pop{}\;
\If{$s \notin hm$}{$hm$.\Insert{$s$}}
$T1$.\Push{$s$}
}
$T = T1$.\Copy{}\;
\Repeat{($T1$.\Size{} $> 1$)}
{
 $t = T1$.\Pop()\;
 $new\_parent =$ \NewParent{}.\AddChild{$t$}\;
 $new\_parent$.\AddinWindow{$t$}\;
  \If{$new\_parent$.\NumChildreninWindow{} $=$ \WindowSize}
  {
     $new\_parent.hash\_id =$ \HF{$new\_parent$.\Children{}}
     \eIf{($new\_parent.hash\_id$ matches $R$) $\And$ ($new\_parent.hash\_id \notin hm$)}
         {$T2$.\Push{$new\_parent$}}
         {$new\_parent$.\RollWindow{}}
  }
 \If{$T1$.\Size{} $== 0$}
    {$T2$.\Push{$new\_parent$}}
 $T = T2.\Copy{}$\; 
 $T1 = T2$\;
 }
return $T$\;



\caption{The \CDMT{} Build Algorithm}
\label{algo:cdmt_build}
\end{algorithm}

We would like to emphasize that the content-defined rule can be any string but should be preferably content-related~\cite{Xia2016FastCDCAF}. A common pattern is to look at last $k$ bits of the hash of the concatenated hash value being zero. In the worst case, if we assume the algorithm detects this pattern for every data chunk, then the number of internal nodes are same as data chunks. Here, the algorithm will be in an infinite loop because it generates an internal node for every leaf node, and the stack size does not reduce. However, for such a condition to arise, the hash function must be heavily skewed in the distribution of the range. If we assume a  hash function with good pseudo-randomness, on average, such a pattern in which last $k$ bits of a hash are all 0 will only exist in  $(1/(2^k))$ data chunks. Therefore, eventually the number of internal nodes in each level would be less than the total number of data chunks and we will find the root node. The complexity of the algorithm is $O(N)$, assuming that 1 parent node is created for each 4 nodes ($N + (1/4)N + (1/16)N + ... ) = (4/3)N$.




\section{\CDMT{} Tree in Containers}
\label{sec:system}


We focus on storage organization of Docker images as Docker is the most popular choice of container engines. We consider this storage from the semantics of user operations and the internal organization. Docker images contain all information needed to run the packaged software~\cite{merkel2014docker}. Images are themselves divided into a series of layers. Every layer, identified by a hash identifier, is a set of files. Docker commands such as push and pull operate at the granularity of images. Thus a user can pull or push from a central server or registry for the \emph{ubuntu:latest} image. Internally,  Docker engine operates at the granularity of layers. Thus given an image, it checks if any of its constituent layer is already present. If any of the layers is present, it does not pull that layer. Docker represents layers as gzip-compressed tar files over the network and on the registry machines.  

Layers in a docker image are read only except for a write-able layer at the top. Modifications or changes made to files in existing layers are stored in the write-able layer. Changes to the write-able layer must be committed and pushed to a registry, which creates a new tagged version of an image. Docker also supports the branching concept where two images may share the same set of parent layer and their ancestors. While a client may make several commits, the Docker registry only recognizes commits that are `tagged’; only `tagged’ images can be pushed or pulled from the hub, and not the intermediate versions.  

Internally, Docker supports different storage drivers; drivers represent layer data in different ways. In this paper, we focus on drivers which store layer data by deduplicating it at subfile granularity (such as Btrfs). By default, Docker chooses the AUFS driver, which uses OverlayFS to store data. OverlayFS is a union file system, which does not store data directly on disk, but uses another file system (e.g., ext4) as underlying storage. Since union file systems support COW (copy-on-write) at file granularity \textit{i.e.}, the entire file is duplicated, and has been shown to have several performance problems~\cite{wu2015totalcow}~\cite{harter2016slacker}, we do not consider the AUFS storage driver. 

Our prototype implementation of the Btrfs driver consists of a deduplicated storage with three components: (i) a container store, which stores unique CDC-defined chunks in a log-structured storage, (ii) a fingerprint index which records fingerprints of a chunk and also maps fingerprints of chunks to their physical locations. This index is used to identify duplicate chunks, and (iii) a recipe store that stores in list format the logical fingerprint sequence of each layer. A recipe is used to reconstruct a layer during restore. 

In our prototype \CDMT{} is the fingerprint index, which determines which chunks to store in the deduplicated storage. Traditional deduplication storage systems implement the fingerprint index as a key-value index in which the key is a fingerprint and the value points to the chunk.

Due to fingerprints being completely random, such an index exhibits poor random access performance for lookup operations\footnote{Deduplication systems, such as~\cite{zhang2016fast} use a cache to speed up the index but caches suffer from the same chunk shift problem addressed in this section}. 

We assume a container registry or server, which hosts all versions of an image along with one \CDMT{} index per image type. We assume the deduplicated storage is resident on a client, which supports push and pull interfaces. The push and pull operations specify images to be pulled or pushed, but operate at the granularity of chunks and use a \CDMT{} index within the client deduplicated storage to determine which chunks to push and pull. The pull and push operations on the client use the \CDMT{} index to determine specific chunks as follows:



\subsubsection{Pull} In the pull operation, the client specifies an image version to the server, and the server sends the \CDMT{} index for the specific image version, which the client compares with \CDMT{} index on the client, and based on the index comparison the client pulls specific blocks for an image. To compare the pull operation performs a breadth-first search over the two \CDMT{} indexes. The input variable root is the root node on the client side. The \CDMT{} index it compares with is the \CDMT{} index of the specific image version the client wants to pull, and hash map is assumed to have the identifier for each version for all the nodes. Algorithm~\ref{algo:cdmt_compare} describes the comparison.

\subsubsection{Push} 
Push occurs in two situations: the push of a new image and push of an already committed image. In the pushing a new image, the client pushes all the chunks to the server along with its specific \CDMT{} index. For pushing an image which the client has already committed, the client requests the \CDMT{} index of the last version from the server, and compares it with the client \CDMT{} index to determine the chunks that have changed. Algorithm~\ref{algo:cdmt_compare} compares the two indices. It sends to the server the chunks corresponding to the image, the layers to which they belong, and the new \CDMT{} index. The server maintains a single \CDMT{} index for different images versions as described in the next section.

 \begin{algorithm}
 \SetAlgoLined
 \DontPrintSemicolon
 \SetKwInOut{Input}{Input}\SetKwInOut{Output}{Output}
 \Input{hash map $hm$, root node of $T1$ and $T2$, versions $v$} 
 \Output{chunks that are different between $T1$ and $T2$}
 \BlankLine
 create a FIFO Queue \textit{queue} of $T1$ and $T2$\;
 push \textit{root} into \textit{queues}\;
 \While{\textit{queue} is not empty}{
      node \textit{current} = Queue.pop(\textit{queue})
     \If{\textit{current.fp} is not in the version \textit{v}}{
         \eIf{\textit{current} has children}
         {
              push all the children of the \textit{current} into \textit{queue}
         }
        {
             yield \textit{current}
        }
        }
        }
 \caption{\CDMT{} compare}
 \label{algo:cdmt_compare}
 
 \end{algorithm}

\subsection{Maintaining the \CDMT{} Index}
\begin{figure}[h]
\centering
\includegraphics[width=2.5in]{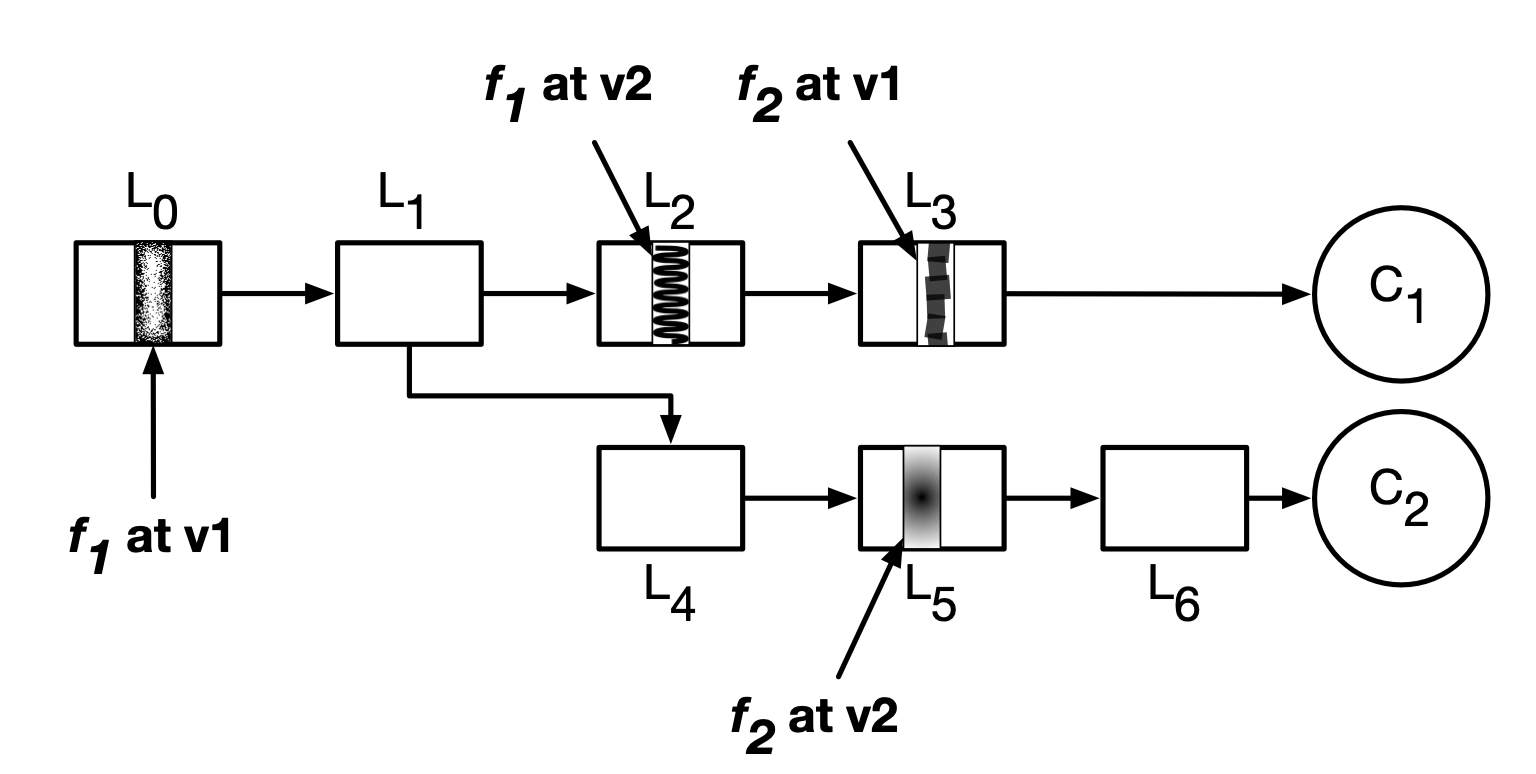}
\caption{The layered filesystem in containers. Containers $C_1$ uses files $f_1$@v2 and $f_2$@v1; container $C_2$ uses file $f_1$@v1 and $f_2$@v2}
\label{fig:layers}
\end{figure}

The client pulls a container image, updates the image layers, and pushes a new version of the image to the server. Most container engines on a client use copy-on-write (COW) such that a block is never overwritten but simply versioned. While maintaining \CDMT{}, we distinguish the versioning of a block due to COW and versioning of a block due to a push, which creates a new branch or tagged version on the server. We illustrate this through an example (Figure~\ref{fig:layers}). Figure shows two container images, $C_1$ and $C_2$, in which one container is a branch of the other, or simply shares layers. $C_1$ is an image consisting of four layers ($L_0$-$L_3$), and $C_2$ consists of five layers ($L_0$-$L_1$,$L_4$-$L_6$). As shown, the file $f_1$, but different versions, resides in layers $L_0$ and $L_2$, and the  file $f_2$, but different versions, resides in the two different branches. The container $C_1$  only
sees the latest version of $f_1$ from $L_2$ due to COW. However, $C_1$ and $C_3$ belong to two different branches and each container sees the most recent version of the file $f_2$ in its respective branch. 


The client within the deduplicated storage determines the source of versioning, \textit{i.e.}, if versioning occurs due to COW or versioning occurs due to branching. Versioning due to COW overrides old dependencies with new dependencies (for \emph{e.g.}, one layer using \emph{gcc6.1} and the other used \emph{gcc7.2}; these new versions are informed by COW. Versioning due to branching is a user-defined operation (for \textit{e.g.}, modifications in example Figures~\ref{fig:MTshift} and \ref{fig:CDMTshift}); the system determined these versions because of explicit push/pull calls. 

\begin{figure}[h]
\centering
\includegraphics[width=3in]{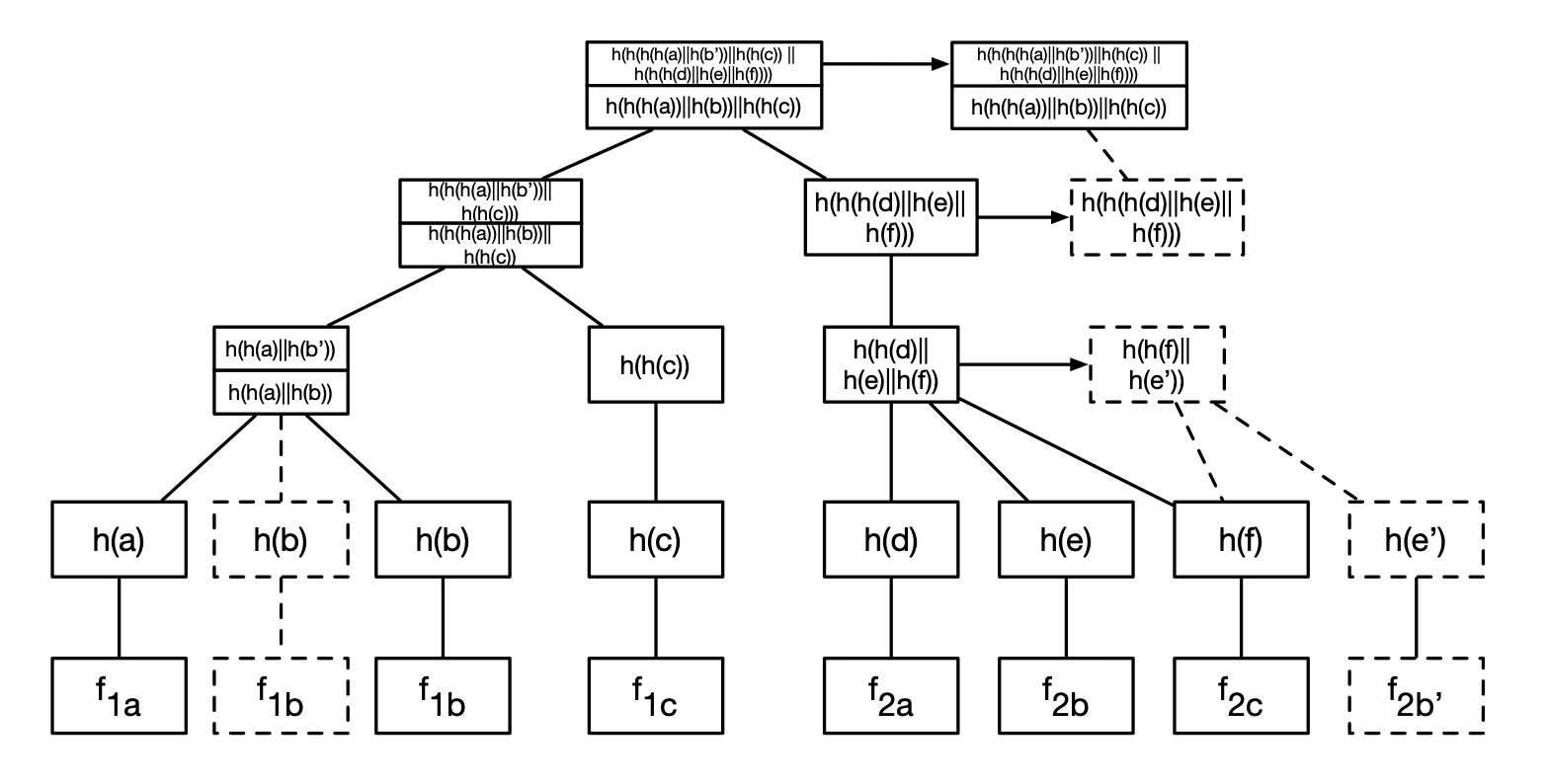}
\caption{Versioning due to layering for $f_1$ and versioning due to branching for $f_2$. Layering leads to new hash versions in internal nodes. Branching leads to new hashing nodes.}
\label{fig:versioning}
\end{figure}

We illustrate the two types of versioning in the \CDMT{} index. Figure~\ref{fig:versioning} shows a versioning due to layering, and versioning due to branching both maintained as part of the same \CDMT{} index. 

To maintain versioning due to layering, we  add a modification history to every internal node of the \CDMT{} index. The left most branch of the \CDMT{} index in Figure~\ref{fig:versioning} depicts this modification history. 
Thus, each internal node knows what its value was at any previous point in time. This  technique of storing versions within an internal node requires space equal to the number of hash bits for every modification. 
However, we must find the right version at each node as we traverse the structure, and this takes time proportional to the number of modifications. So if $m$ modifications are made, then each access operation
has O(log $m$) slowdown. In most cases, clients work with the most recent version However, since we only have to look at the most current version every time due to layering the expensive access time is not incurred unless old versions are heavily accessed. 

To maintain versioning due to branching, we add a modification history to the \CDMT{} index. For this we make a copy of any \CDMT{} node before changing it, and we cascade the change back through the data structure, \textit{i.e.}, all nodes that pointed to the old node are modified to point to the new node instead. These modifications cause more cascading changes, and so on, until the root. We maintain the root as an array of roots where each root corresponds to a `taggable` container branch, and the nodes pointed by a specific version of root are the nodes corresponding to that specific branch. Maintaining versioning due to branches only adds O(log $m$) additive lookup time, where $m$ is the number of modifications, and amounts to finding the specific branch.

%

\section{Experiments}
\label{sec:experiments}

Our experiment dataset consists of container images from 15 different application across four categories. The categories include programming languages, web frameworks and servers, databases and data science. We base our choice of images on applications which are being actively developed and thus have a large number of versions so that we can assess the impact of push/pull operations during container delivery. Table~\ref{tab:dataset} shows the 15 applications, the number of versions considered for each image, and the average number of layers per application. Finally, the Table shows the total size of an application across all versions (layers are duplicated). The images were downloaded from Docker Hub between 8$^{th}$ and 14$^{th}$ of July, 2020 (for \emph{golang}, \emph{node}, \emph{tomcat}, \emph{python}, \emph{tensorflow}, \emph{R-studio/r-base}, \emph{radis}, and \emph{rails}), and  between 10$^{th}$ and 20$^{th}$ October 2020 (for \emph{httpd}, \emph{nginx}, \emph{postgres}, \emph{django}, \emph{pytorch}, and \emph{deepmind}).


We perform all experiments on a machine running Linux kernel 2.6.32-754.33.1.el6.x86\_64 CentOS 6.10, having Intel(R) Xeon(R) CPU E5-2623 v3 @ 3.00GHz, 3 terabytes of hard drive and 64 GB of internal memory. We perform some preprocessing on image data before conducting our experiments. Image versions are uncompressed for deduplication as the specific \textit{tar} archive format changes the order of files and inserts file headers. In CDC our window size is 2 and for Merkle trees our chosen $k$ is 4.

\begin{table}[]
\caption{\small The dataset: 15 applications sourced from Docker Hub.}
\begin{tabular}{|c|c|c|c|}
\hline
\textbf{Application} & \textbf{\# of versions} & \textbf{\begin{tabular}[c]{@{}c@{}}Avg. \# of layers \\ per version\end{tabular}} & \textbf{Total size(Gb)} \\ \hline
golang  & 8 & 5.3 & 2.5            \\ \hline
node    & 17 & 3.2 & 1.3            \\ \hline
tomcat  & 17 & 6.3 & 3.2            \\ \hline
httpd   & 17 & 5.0 & 2.0            \\ \hline
python  & 18 & 4.9 & 1.7            \\ \hline
tensorflow  & 10 & 24 & 24             \\ \hline
r-base  & 9 & 8 & 35             \\ \hline
redis   & 13 & 6 & 0.83           \\ \hline
rails   & 18 & 17 & 53             \\ \hline
nginx   & 34 & 3.4 & 1.1            \\ \hline
postgres    & 19 & 8.9 & 1.1            \\ \hline
django  & 8 & 8 & 4.2            \\ \hline
pytorch & 10 & 7.9 & 89             \\ \hline
mysql   & 16 & 12 & 7.4            \\ \hline
deepmind    & 19 & 15 & 100            \\ \hline
\textbf{Total}  & 233 & 8.4 & 320            \\ 
\hline
\end{tabular}
\label{tab:dataset}
\end{table}


\subsection{Block-based deduplication reduces storage size}
\begin{figure}
    \centering
    \includegraphics[width=\linewidth]{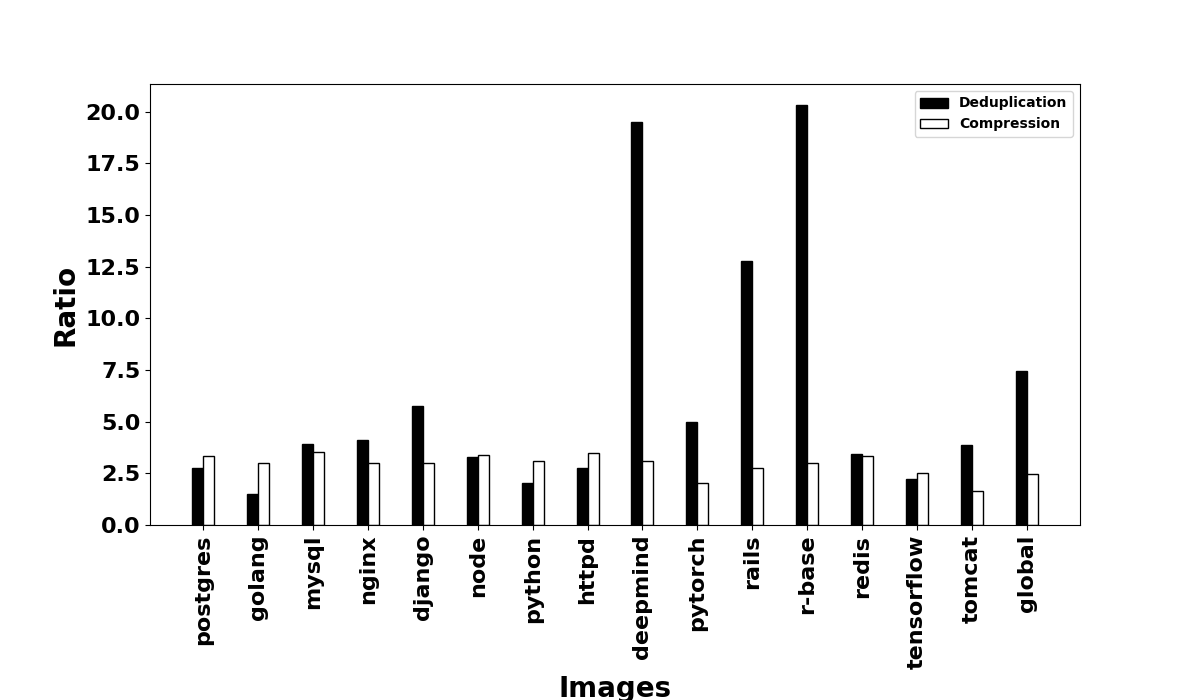}
    \caption{The performance of compression ratio by deduplication and gzip compression algorithm on individual images.}
    \label{fig:compression-ratios}
\end{figure}

\begin{figure}
    \centering
    \includegraphics[width=\linewidth]{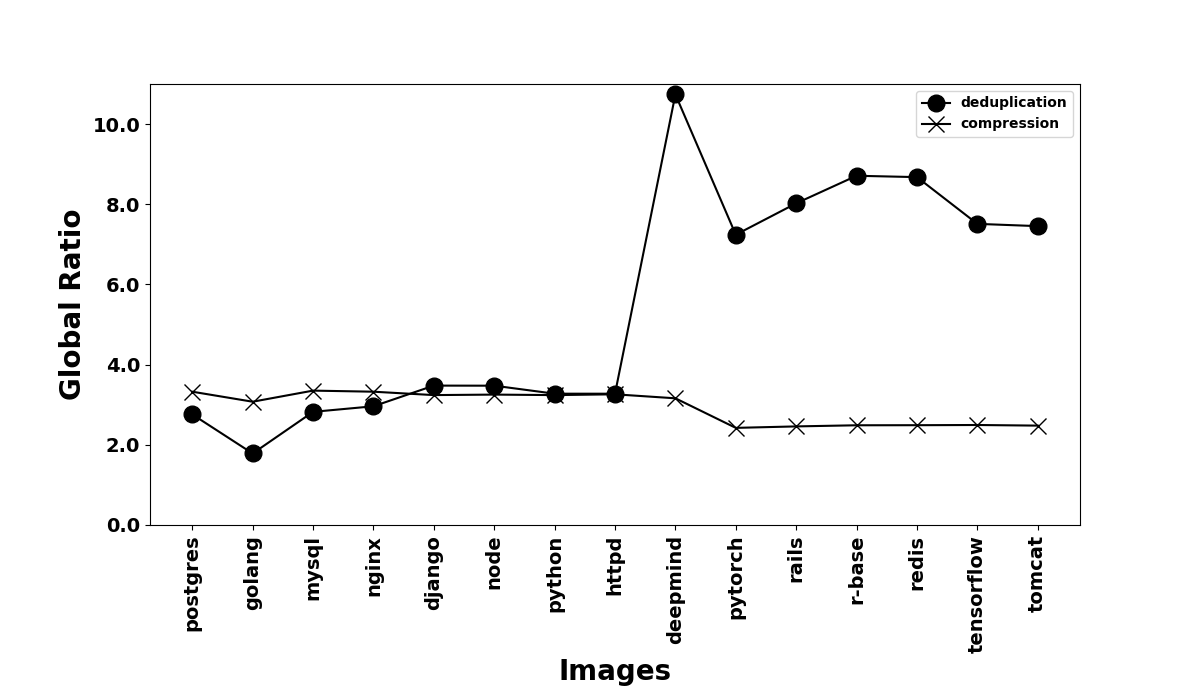}
    \caption{The performance of global deduplication and gzip compression as the dataset increases.}
    \label{fig:progression-compression-ratios}
\end{figure}

Using the 233 image versions, we show that block-based deduplication reduces total storage size. Figure~\ref{fig:compression-ratios} shows the ratios of the size of the raw data to the size of the compressed or deduplicated data. We average the compression ratio over all versions for each software, and similarly compute the deduplication ratio over all versions. Note the higher the value, the more reduction in size. Compression ratios only reach as high as 3.5 but deduplication ratio reach as high as 20 for \textit{deepmind}. Deduplication is not better for every image, though for more than half the number of images deduplication performs better than gzip compression. T. Harter \textit{et.} \textit{al.}~\cite{harter2016slacker} choose compression as better for a single version, but as we show deduplication is better over multiple versions and is an important consideration for container registries. For clients, we consider global deduplication where images are deduplicated by aggregating different applications together. Figure~\ref{fig:progression-compression-ratios} captures the performance of global block deduplication as the number of images in the dataset grow. Figure shows global deduplication ratio to be about 7.7 when compression ratio is about 2.5, showing the advantage of block-based deduplication also in clients. 

\subsection{The \CDMT{} index reduces network and disk I/O} 

\begin{figure}
    \centering
    \includegraphics[width=\linewidth]{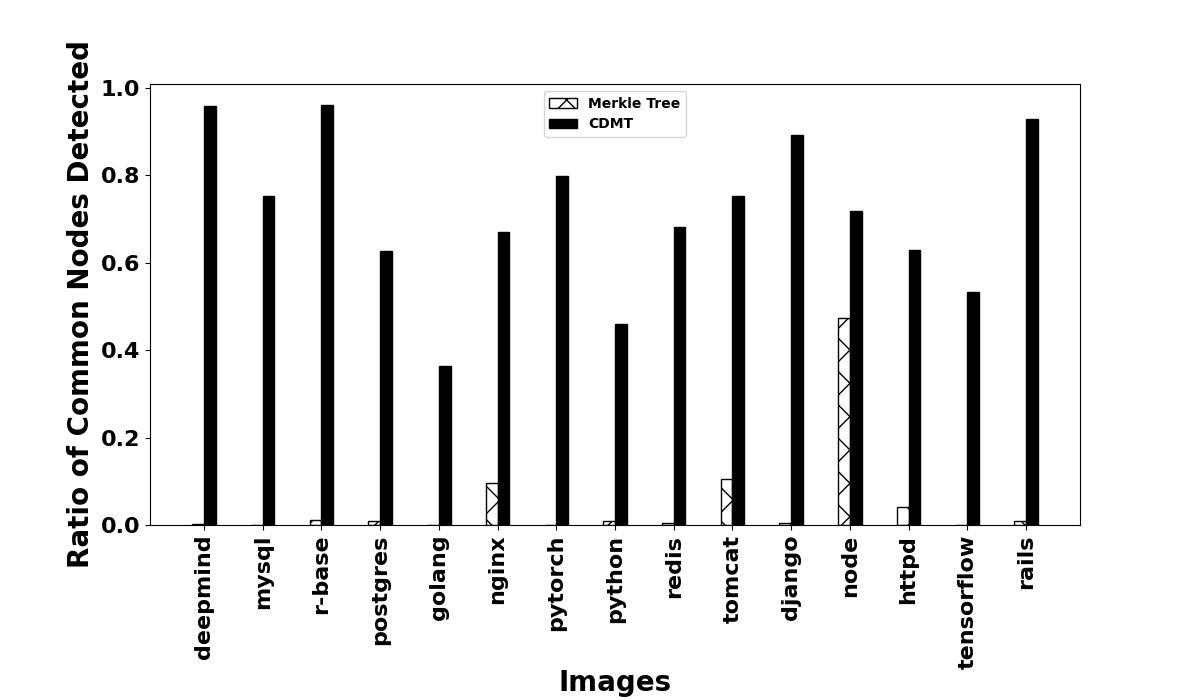}
    \caption{The comparison of the \CDMT{} index and Merkle Trees generated for the same versions, averaged over all images. \CDMT{} indexes for two versions have higher ratio of common nodes detected because nodes are not cut due to the chunk shift problem as in Merkle Trees.}
    \label{fig:common-nodes-ratio}
\end{figure}

We compare the advantage of using \CDMT{} index over Merkle trees in block-based deduplication storage. Figure~\ref{fig:common-nodes-ratio} compares the \CDMT{} index with with Merkle trees and shows that \CDMT{} indexes detect common data blocks significantly more than Merkle trees.  Merkle Tree has a very low number of common nodes except for \textit{nginx, tomcat, node}. This observation shows chunk-shift happens fairly often, if not all the time, making Merkle Tree inefficient as an index over CDC-defined chunks.

We translate this result further in terms of push and pull operations on a container. In particular, we consider a client which pulls different versions of an application and upgrades it. Table~\ref{tab:my_label} reports the block-based average deduplication ratio, i.e., the average number of common blocks determined using \CDMT{} tree (column 1) and determines how many blocks the client pulls in terms of data size (column 2).  The second column shows the amount of network traffic that will occur if the client identifies the correct number of deduplicated blocks, and the client pulls only the non-deduplicated blocks over the network. As we show identifying common blocks Figure~\ref{fig:common-nodes-ratio} is possible through \CDMT{} but not via Merkle tree. The first column of Table~\ref{tab:my_label} when multiplied by the total size of an image indicates the amount of disk space saved.

\begin{table}[]
    \centering
        \caption{\small Amount of disk I/O and network I/O over as new container versions are pulled}
    \begin{tabular}{|c|c|c|}
        \hline
        \textbf{Application} & \begin{tabular}[c]{@{}c@{}} \textbf{Block-based} \\ \textbf{deduplication ratio} \end{tabular} & \begin{tabular}[c]{@{}c@{}} \textbf{Total non-deduplicated} \\ \textbf{size (gb)} \end{tabular}\\
        \hline
        golang & 0.34 & 1.7  \\
        \hline
        node & 0.70 & 0.41 \\
        \hline
        tomcat & 0.74 & 0.88 \\
        \hline
        httpd & 0.64 & 0.73 \\
        \hline
        python & 0.50 & 0.90 \\
        \hline
        tensorflow & 0.55 & 9.8 \\
        \hline
        r-base & 0.95 & 1.8 \\
        \hline
        redis & 0.71 & 0.25 \\
        \hline
        rails & 0.92 & 4.5 \\
        \hline
        nginx & 0.76 & 0.28 \\
        \hline
        postgres & 0.64 & 0.42 \\
        \hline 
        django & 0.83 & 0.76 \\
        \hline
        pytorch & 0.80 & 19 \\
        \hline
        mysql & 0.74 & 2.1 \\
        \hline
        deepmind & 0.95 & 5.3 \\
        \hline
    \end{tabular}

    \label{tab:my_label}
\end{table}

\subsection{Reduction in Comparisons}
\begin{figure}
    \centering
    \includegraphics[width=\linewidth]{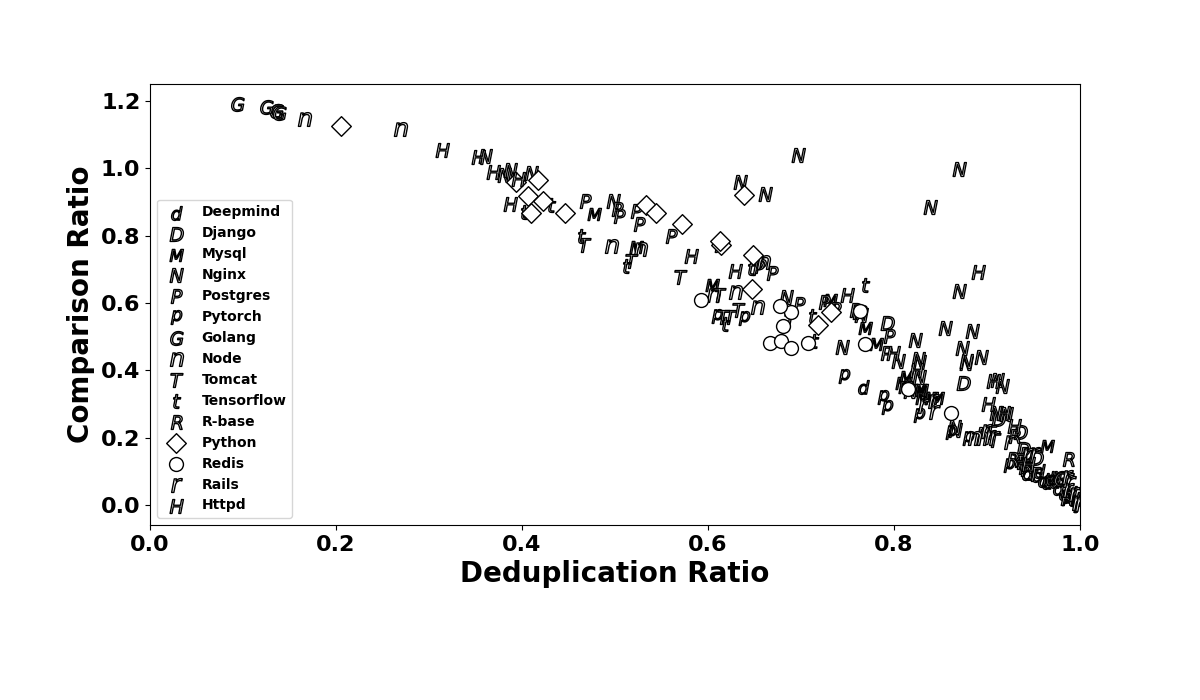}
    \caption{Relationship between comparison and deduplication ratios using \CDMT{}. As versions increase in similarity, the number of comparisons required to identify common data chunks decreases nearly linearly.}
    \label{fig:comparisons-vs-dedup-ratios}
\end{figure}

We further determine if authentication paths in \CDMT{} are useful for determining if a matching node was found. For this experiment, we define a comparison ratio as the number of nodes compared with the \CDMT{} index divided by the number of nodes of compared using simple key-value search for finding matching chunks. When the ratio is over 1 it shows the number of comparisons will be more with \CDMT{} index, \textit{i.e.}, authentication paths are not being useful, and less than 1 shows the number of comparison will be smaller with \CDMT{} index over content-defined chunks, \textit{i.e.}, authentication paths are helpful and we can skip traversing the entire tree below that node in the authentication path. When the ratio is over 1 then key-value search is sufficient. For versions that have high similarity a lower comparison ratio can save a significant number of comparisons.  Figure \ref{fig:comparisons-vs-dedup-ratios} shows this relationship between comparison ratio and deduplication ratio with the \CDMT{} index. We observe that as versions of images become increasingly similar, the number of comparisons to detect matching chunks between them decreases nearly linearly. The result is especially remarkable as the \CDMT{} index is not a balanced tree like Merkle tree whose traversal is known to have sub-linear time complexity \cite{BERMAN200726}.

\subsection{Construction time for \CDMT{}}
\begin{figure}
    \centering
    \includegraphics[width=\linewidth]{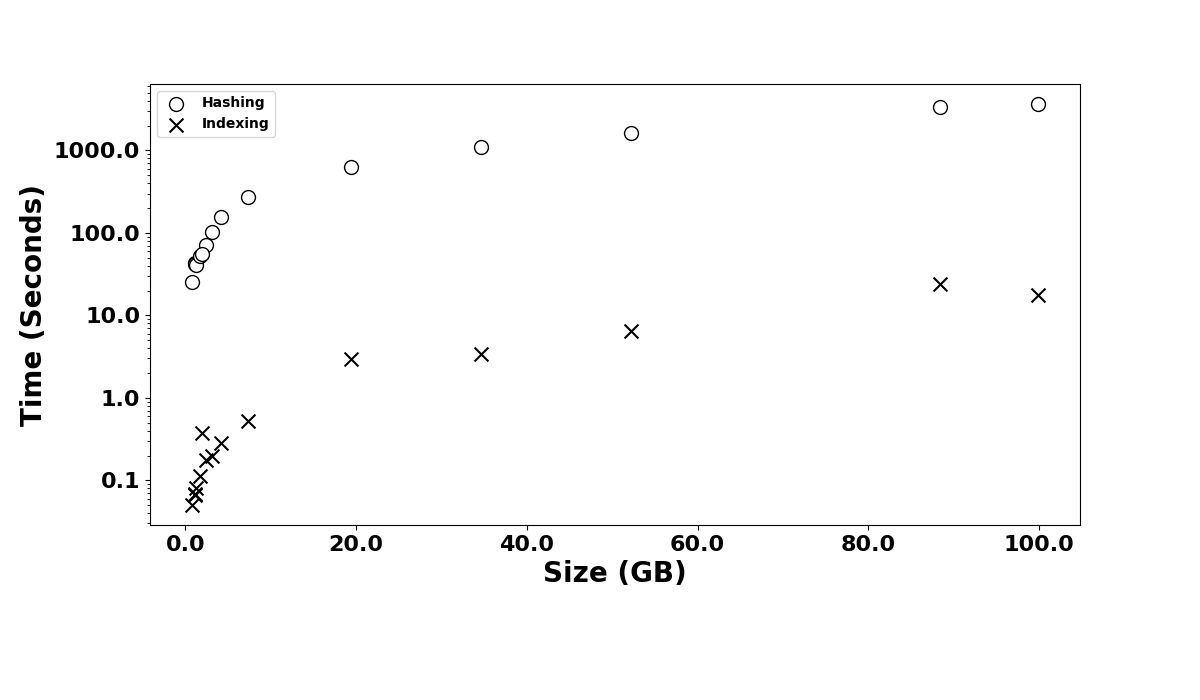}
    \caption{Time taken to construct the \CDMT{} index for the versions of an image is significantly less than time taken to create content-defined hashes for the same image.}
    \label{fig:hashing-vs-tree-time}
\end{figure}

\CDMT{}s are lightweight to construct and transmit over the network. For this experiment, we treat the images for each software in our dataset as versions of each other. We convert each version into its corresponding chunks using our hashing algorithm and then construct the \CDMT{} index. The hashing time is the computation time for both setting the boundary and computing the hash value for each chunk. The indexing time is building the \CDMT{} index per version using Algorithm~\ref{algo:cdmt_build}. We compare hashing versus indexing time in Figure \ref{fig:hashing-vs-tree-time}. In this experiment, content-defined Chunking(CDC) is performed using Rabin fingerprint in which hash value are computed through linear congruence of polynomials. The chunk boundary computation is linear-time over the number of characters in the data. After CDC sets the chunk boundary, Blake2b~\cite{blake2b} is used to compute the hash value of the chunk because it is secure and provides maximum efficiency for large data sets. As the Figure shows the time taken to construct the \CDMT{} index for each image (over all the versions) is only a fraction of the time needed to produce all its hashes. This motivates us to reduce the hashing time, either to change the hash function for setting boundaries, or only hashing a fraction of the nodes. We plan to consider these strategies in future work.

\section{Conclusion}
\label{sec:conclusion}
Containerized applications have become popular, and hosting containers on a local data hub is widespread. Due to isolation of virtualized applications, storage requirements of containers can be high. In this paper, we have argued for block-based deduplication of container storage but also shown that we must support the deduplication layer with a content-defined Merkle tree index to efficiently push/pull layers over the network. We have shown the efficiency of our approach over namespaced containers. The \CDMT{} tree is general and can also apply to aggregations of virtualized images. 

\section*{Acknowledgment}
This work is supported by National Science Foundation under grants CNS-1846418, NSF ICER-1639759, ICER-1661918 and a Department of Energy Fellowship. 

\bibliographystyle{IEEEtran}

\end{document}